\documentclass[12pt,preprint]{aastex}
\usepackage{natbib}
\usepackage{color}
\newcommand{\Ha}{H$\alpha$}
\newcommand{\Hb}{H$\beta$}

\newcommand{\CaII}{Ca\,{\sc ii}}
\newcommand{\CII}{C\,{\sc ii}} 
 
\newcommand{\OI}{w}
\newcommand{\HII}{H\,{\sc ii}}
\newcommand{\SiII}{Si\,{\sc ii}}

\newcommand{\HeI}{He\,{\sc i}}
\newcommand{\FeII}{Fe\,{\sc ii}}
\newcommand{\MgI}{Mg\,{\sc i}}

\newcommand{\NaID}{Na\,{\sc i}~D}

\newcommand{\Mo}{M$_{\odot}$}

\shorttitle{SN~2011bm}
\shortauthors{Valenti et al.}

\begin{document}

\title{A spectroscopically normal type Ic supernova from a very massive progenitor}

\author{Stefano Valenti\altaffilmark{1}; Stefan Taubenberger\altaffilmark{2};
Andrea Pastorello\altaffilmark{1};  Levon Aramyan\altaffilmark{3}; 
Maria Teresa Botticella\altaffilmark{4}; Morgan Fraser\altaffilmark{5};
Stefano Benetti\altaffilmark{1}; Stephen J. Smartt\altaffilmark{5}; 
 Enrico Cappellaro\altaffilmark{1};  Nancy Elias-Rosa\altaffilmark{6}; 
 Mattias Ergon\altaffilmark{7};  Lindsay Magill\altaffilmark{5}; 
 Eugene Magnier\altaffilmark{8}; Rubina Kotak\altaffilmark{5}; 
 Paul A. Price\altaffilmark{9}; Jesper Sollerman\altaffilmark{7}; 
 Lina Tomasella\altaffilmark{1};  Massimo Turatto\altaffilmark{1}; 
 Darryl Edmund Wright\altaffilmark{5}}

\altaffiltext{1}{INAF-Osservatorio Astronomico di Padova, vicolo
dell'Osservatorio 5, 35122 Padova, Italy}

\altaffiltext{2}{Max-Planck-Institut f\"ur Astrophysik,
Karl-Schwarzschild-Str. 1, 85741 Garching bei M\"unchen, Germany}

\altaffiltext{3}{Department of General Physics and Astrophysics, Yerevan State University, 1 Alex Manoogian, 0025 Yerevan, Armenia}

\altaffiltext{4}{ INAF- Osservatorio Astronomico di Capodimonte,
Salita Moiariello, 80128 Napoli, Italy}

\altaffiltext{5}{Astrophysics Research Centre, School of Mathematics and
Physics, Queen's University Belfast, Belfast BT7 1NN, UK}

\altaffiltext{6}{ Institut de Ciències de l'Espai (IEEC-CSIC), Facultat de Ciències, Campus UAB, 08193 Bellaterra, Spain}

\altaffiltext{7}{The Oskar Klein Centre, Department of Astronomy, AlbaNova, Stockholm University, 10691 Stockholm, Sweden}

\altaffiltext{8}{Institute for Astronomy, University of Hawaii, 2680 Woodlawn Drive, Honolulu, HI 96822, USA}

\altaffiltext{9}{Department of Astrophysical Sciences, Princeton University, Princeton, NJ 08544, USA}

\begin{abstract}
We present observations of the Type Ic supernova (SN~Ic) 2011bm
spanning a period of about one year.  The data establish that
SN~2011bm is a spectroscopically normal SN~Ic with moderately low
ejecta velocities and with a very slow spectroscopic and photometric
evolution (more than twice as slow as SN~1998bw).  The Pan-STARRS1
retrospective detection shows that the rise time from explosion to
peak was $\sim$\,40 days in the R band. Through an analysis of the
light curve and the spectral sequence, we estimate a kinetic energy of
$\sim$\,7--17 foe and a total ejected mass of $\sim$\,7--17 \Mo, 5--10
\Mo{} of which is oxygen and 0.6--0.7 \Mo{} is $^{56}$Ni. The physical
parameters obtained for SN~2011bm suggest that its progenitor was a
massive star of initial mass 30-50 \Mo.  The profile of the forbidden
oxygen lines in the nebular spectra show no evidence of a bi-polar
geometry in the ejected material.
\end{abstract}


\section{Introduction}
The standard classification scheme for the explosions of massive stars
consists of two different branches: stars retaining their hydrogen
envelope at the time of the explosion produce Type II supernovae
(SNe~II), whilst those losing it before the explosion produce Type Ib
or Ic SNe, depending on the strength of He lines in the spectra. In
addition, some SNe~Ic, labelled BL-Ic, show broad lines in the
spectra, signature of a very high kinetic energy/ejected mass
ratio. Over the past 5 years, the above scenario has changed
dramatically with the discovery of new classes of transients that may
originate from different explosion channels. Pair-instability,
pulsational pair-instability and magnetar-powered explosions have been
suggested to explain the properties of a group of hyper-luminous and
slowly evolving transients (SN~2007bi, \citealt{2009Natur.462..624G};
SN~2006gy, \citealt{2007ApJ...666.1116S}; SN~2010gx,
\citealt{2010ApJ...724L..16P}, \citealt{2011Natur.474..487Q}), while
fall-back on the collapsed remnant, electron-capture and
failed-deflagration scenarios have been proposed to explain faint and
fast-evolving transients (e.g. SN~2008ha,
\citealt{2009Natur.459..674V}, \citealt{2009AJ....138..376F};
SN~2005E, \citealt{2010Natur.465..322P}). Alternatively, some of these
events can still be explained if a more \emph{canonical} core-collapse
scenario but an extended range of physical conditions of the
progenitor at the moment of its explosion is invoked (e.g. SN~2006gy,
\citealt{2009ApJ...691.1348A}; SN~2007bi,
\citealt{2010A&A...512A..70Y,2010ApJ...717L..83M}; SN~2005cz,
\citealt{2010Natur.465..326K}). Here we present the observations of a
stripped-envelope supernova that evolved in an extremely slow fashion,
much more slowly than any other \emph{spectroscopically normal}
core-collapse SN studied in the literature.

\section{Observations}
\begin{figure*}
\plotone{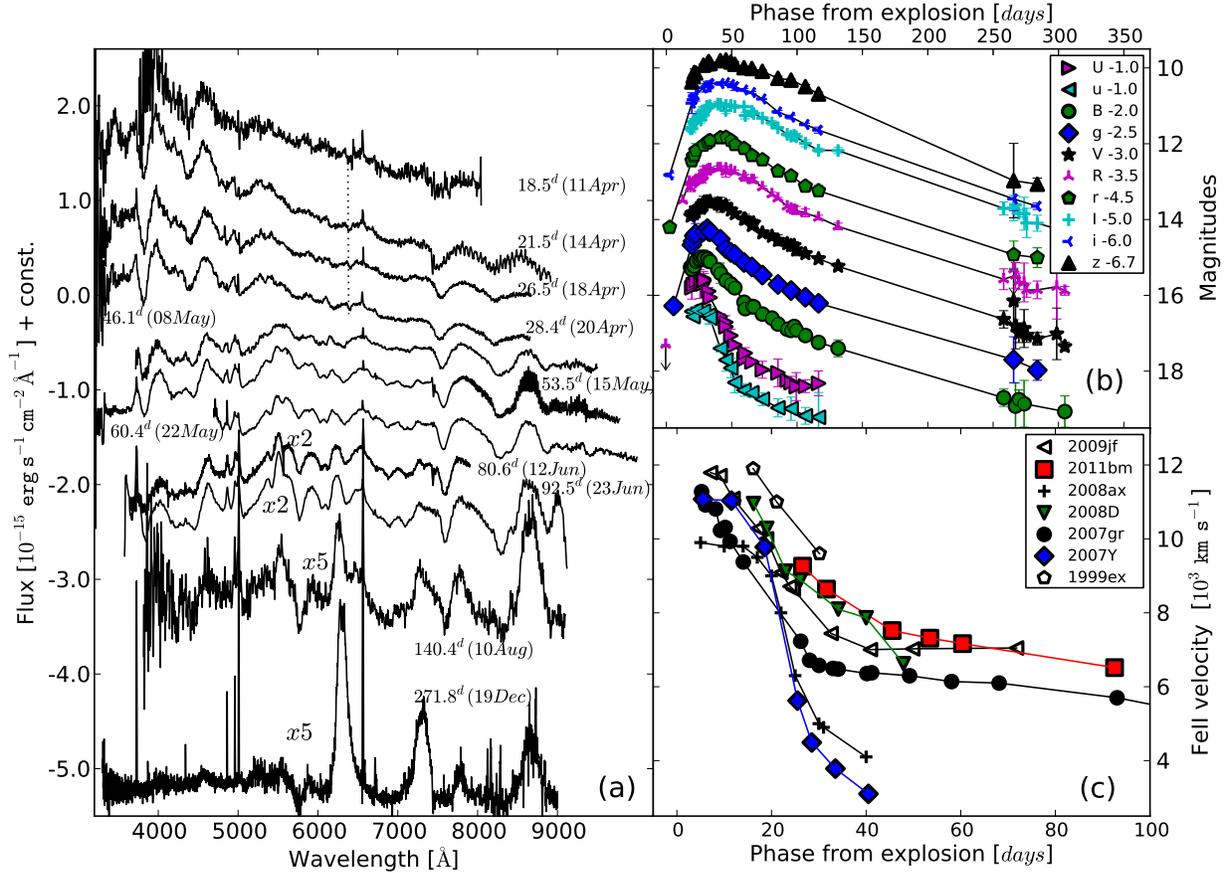}
\caption{Spectroscopic and photometric follow-up of SN~2011bm. All
  data have been reduced with standard IRAF routines, using the QUBA
  pipeline (see \citealt{2011MNRAS.416.3138V} for information). For
  the NOT spectra, a second order correction was performed using the
  method described in \protect\citet{2007AN....328..948S}.
  \textbf{a}: Sub-sample of our spectra in the host-galaxy frame. The
  dashed line marks the position of the narrow absorption line at
  $\sim$\,6390\,\AA{} visible in our earliest four spectra.
  \textbf{b}: Multi-band light curves of SN~2011bm. \textbf{c}:
  Photospheric velocities (measured from \FeII\ $\lambda$4924,
  $\lambda$5018 and $\lambda$5169) for a sample of stripped-envelope
  core-collapse SNe. Data sources: SN~2007gr,
  \citet{2009A&A...508..371H}; SN~2008D, \citet{2008Sci...321.1185M};
  SN~2007Y, \citet{2009ApJ...696..713S}; SN~2008ax,
  \citet{2011MNRAS.413.2140T}; SN~2009jf, \citet{2011MNRAS.416.3138V};
  SN~1999ex, \citet{2002AJ....124.2100S}.}
\label{fig1}
\end{figure*}

SN~2011bm was discovered by the ``La Sagra Sky Survey'' on 2011 April
5 in the galaxy IC 3918\footnote{The following parameters have been
  used for IC 3918 in this paper: $\mu$=34.90 mag (radial velocity
  corrected for in-fall onto Virgo of 2799 $km$ $s^{− 1}$ and a Hubble
  constant of 72 $km$ $s^{- 1}$ $Mpc^{- 1}$.), $z$=0.0221 (from narrow
  emission lines of the host galaxy), $E(B-V)_\mathrm{IC~3918}$=0.032
  mag (assuming a similar dust-to-gas ratio in IC~3918 and in the
  Milky Way and comparing the equivalent width of the narrow \NaID\/
  absorption lines from IC~3918 with that from the Milky Way),
  $E(B-V)_\mathrm{Gal}$=0.032 mag
  \protect\citep{1998ApJ...500..525S}.} \citep{2011CBET.2695....1R}
and classified on April 11 at the 1.82-m Mt.\,Ekar Copernico Telescope
as a type Ic SN close to maximum light
\citep{2011CBET.2695....2R}. The Palomar Transient Factory (PTF)
claimed an independent discovery of SN~2011bm on 2011 March 29 and a
stringent pre-discovery limit on March 23 down to 20.8 mag in the R
band \citep{2011ATel.3288....1G}. Since the Panoramic Survey Telescope
and Rapid Response System-1 (PS1) 3$\pi$ survey has already been very
useful to constrain the explosion epoch of other nearby SNe
\citep[e.g. SN~2010ay, ][]{2011arXiv1110.2363S}, we retrospectively
inspected PS1 data and detected SN 2011bm in $r_{P1}$- and
$i_{P1}$-band images on March 26.5 UT at magnitudes $r$=18.71 mag and
$i$=18.82 mag, and in a $g_{P1}$-band image on March 29.6 UT at a
magnitude $g$=18.78 mag. This is the earliest detection of the
supernova, and strongly constrains the explosion to have occurred
between 2011 March 23 and March 26.  After announcement, we
immediately started a follow-up campaign in the framework of the NTT
European Large Programme (ELP) collaboration and we extensively
monitored SN~2011bm using the telescopes available to us. We collected
a large amount of data in the optical domain, complemented by
near-infrared data, especially useful to investigate the presence of
He in the SN ejecta (cf. Tables~\ref{tab1} and \ref{tab2}). A sub-set
of the spectroscopic and photometric data collected by the NTT ELP
collaboration are shown in Figs.~\ref{fig1}a and \ref{fig1}b,
respectively. In Fig.~\ref{fig1}c the expansion velocity as derived
from the position of the minima of \FeII{} lines\footnote{Average of
  the velocities obtained for the lines at $\lambda$4924,
  $\lambda$5018 and $\lambda$5169.} is shown, and compared with those
of normal SNe~Ib/c. SN~2011bm shows typical velocities, but a slow
velocity evolution similar to those of SNe 2009jf or 2007gr.

The classification spectrum taken on April 11 (18 days after the
explosion) is rather blue ($B-V$ $\sim$ 0.35 mag). Such a blue color
is unusual for SN~Ic spectra at this epoch. Early spectra of SN~2011bm
show the classical type Ib/c SN features: \FeII, \OI, \CaII{} and a
faint P-Cygni absorption at $\sim$\,6100 \AA, usually identified as
\SiII.  A narrow line is visible at $\sim$\,6530\,\AA{}
($\sim$\,6390\,\AA{} in the host galaxy rest frame) in the first four
spectra (see Fig.~\ref{fig1}a; marked with a vertical dotted line). A
similar feature was also observed in very early spectra of SN~2007gr
up to few days before maximum, and was identified as \CII{}
$\lambda$6580, with \Ha{} or \HeI{} $\lambda$6678 as possible
alternatives. Using the spectrum synthesis code
SYNOW\footnote{http:/$\!$/www.nhn.ou.edu/\~{}parrent/synow.html}, we
confirm that \CII{} is a consistent identification for this
feature. This supports the idea that our 18-day spectrum of SN~2011bm
is similar to those of younger, canonical SNe Ic. Prominent features
of \HeI{} or H are not detected in the optical domain, and there is no
evidence for the \HeI{} line at $\sim$ 2 micron in our near-infrared
SOFI spectrum obtained on 2011 May 9, consistent with the
classification as a SN Ic.

\begin{figure*}
\hspace{0.5in}
\epsscale{1}
\plotone{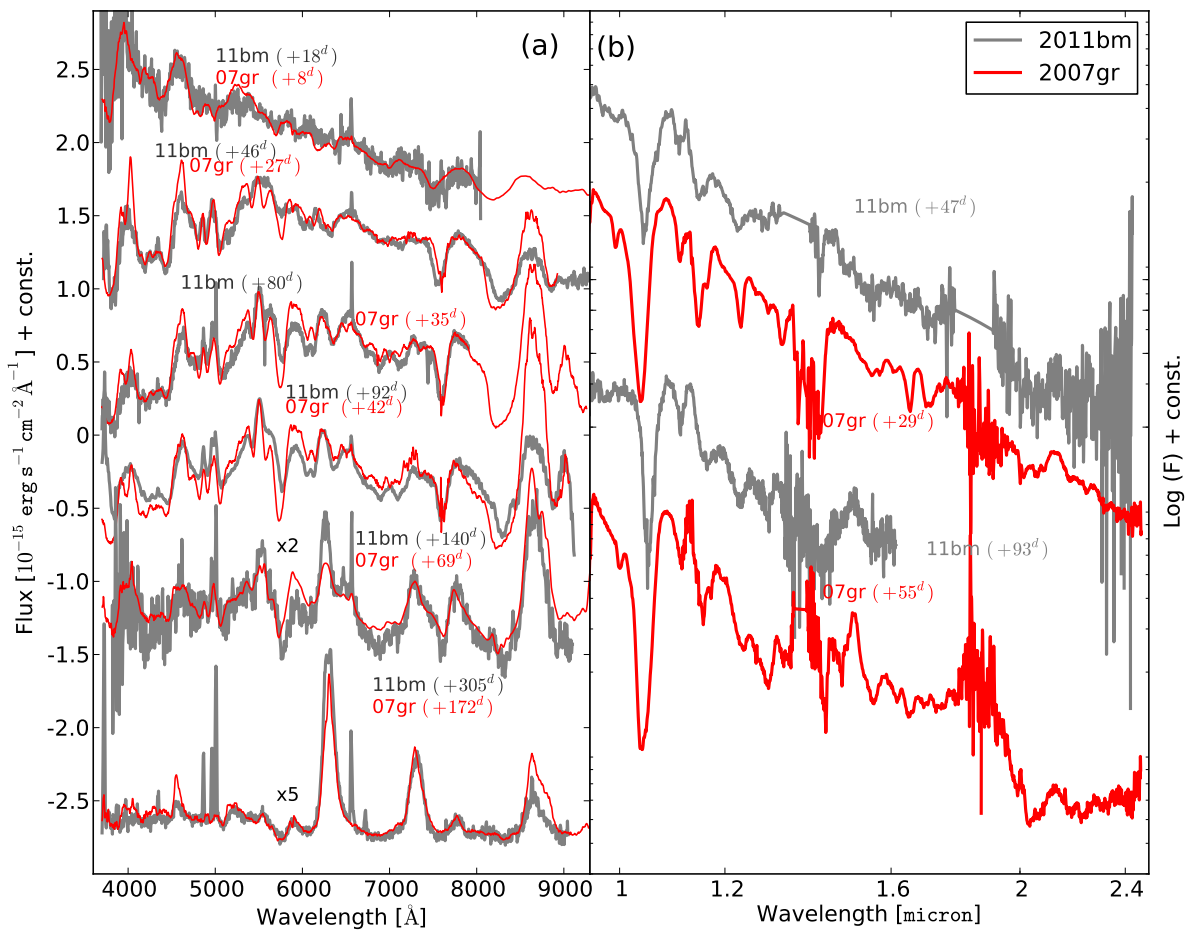}
\caption{\textbf{a}: Spectral evolution of SN~2011bm, and comparison
  with spectra of SN~2007gr. The spectra of SN~2007gr have been scaled
  to match those of SN~2011bm. \textbf{b}: Infrared spectra of
  SN~2011bm compared with spectra of SN~2007gr.}
\label{fig2}
\end{figure*}

The best match for the spectra of SN 2011bm is found with those of
SN~2007gr \citep{2009A&A...508..371H}, persisting along the whole
evolution (Fig.~\ref{fig2}a and \ref{fig2}b). What is different,
however, is the time scale of the transition from optically thick to
optically thin ejecta.  At 18 days past core-collapse the spectrum of
SN~2011bm is still blue (like those of normal type Ic SNe a few days
after the explosion). Three months after the explosion SN~2011bm is
still optically thick, and the best match is found with spectra of
SN~2007gr at a phase of only 1 month. Later on, at 305 days the
spectrum of SN~2011bm shows nebular lines, though the detection of
[\OI] $\lambda$5577 is unexpected, as this line usually disappears in
the spectra of SNe~Ic by $\sim$\,150-–200 days after core-collapse.
The spectrum of SN 2007gr at 172 days shows a stronger \MgI]
  $\lambda$4571 line than that of SN 2011bm. This may be due to a
  lower abundance of magnesium in SN 2011bm than in SN~2007gr or to a
  different ionisation/excitation state of the ejecta in the two SNe
  or simply to the fact that the O-Ne-Mg layer is still not completely
  exposed, while we are still seeing the C/O-rich layer. In the latter
  case, the \MgI] $\lambda$4571 line is expected to become more
    prominent with time.

Even more intriguing is the luminosity evolution of SN~2011bm. After
the earliest detection by PS1, the light curve rises for several
weeks, reaching its maximum in the R-band on 2011 May 2 (JD =
2455684.3), 40 days after the explosion (see Fig. \ref{fig1}b).  No
other \emph{normal} SN Ib/c has shown such a slowly rising light
curve. The post-maximum evolution is qualitatively similar to those of
other SNe~Ic, with a more rapid initial luminosity drop followed by a
slower decline.  However, for SN~2011bm the inflection points occurs
much later than in other SNe Ic. Comparing the light curves of
SN~2011bm with those of SNe~1998bw \citep[R band,
][]{2001ApJ...555..900P} and 1997ef \citep[V band,
][]{2000ApJ...534..660I}, which are among the most slowly evolving
BL-Ic SNe available in the literature, we find that SN~2011bm is
evolving $\sim$\,1.5 times more slowly than SN~1997ef and $\sim$\,2.2
times more slowly than SN~1998bw\footnote{The time evolution is
  computed stretching the light curves around maximum until the light
  curves shapes match.}.  SN 2011bm also evolves 2.7 times more slowly
that SN~2007gr \citep{2009A&A...508..371H} in the R band, in good
agreement with the different time scales of the spectral evolution
highlighted in Fig.~\ref{fig2}.

\section{Physical Parameters}
A direct estimate of the host-galaxy metallicity can be obtained from
the fluxes of nebular emission lines in the vicinity of the SN
location. We measured the following narrow lines of the host-galaxy
detected in our WHT spectrum of 2011 December 19: [O\,{\sc ii}]
$\lambda\lambda$3726,3729, \Hb{}, [O\,{\sc iii}] $\lambda$4959,
[O\,{\sc iii}] $\lambda$5007, \Ha{}, [N\,{\sc ii}] $\lambda$6583.
Using different metallicity indicators, we obtain the following
values: the N2 index calibration of \citet{2004MNRAS.348L..59P} gives
12+log(O/H) = 8.41 dex, while the O3N2 index provided by the same
authors gives a value of 8.31 dex. The method adopted by
\citet{2004ApJ...617..240K} (see their eq.~18) yields 12+log(O/H) =
8.58 dex\footnote{With the caveat that this value may be slightly
  dependent on the used value for the galactic extinction.}. These
values are only marginally sub-solar.  From the \Ha{} emission we also
estimate a star formation rate of $\sim$ 0.06 \Mo{} $yr^{-1}$ that is
typical of \HII{} regions and similar to that of other stripped
envelope SNe \citep{2008MNRAS.383.1485V}.  Our observations of
SN~2011bm can be used to understand the physical conditions of the
progenitor at the moment of the explosion. Important insights come
from the nebular spectra, the shape of the light curve and the
bolometric luminosity.  The spectrum obtained on 2011 December 18
($\sim$\,270 days after the explosion) provides additional information
on the progenitor's properties. The presence of the [\OI]
$\lambda$5577 feature is remarkable: this line is usually visible only
up to $\sim$\,150--200 days after the explosion, because it requires a
relatively high electron density. Using eq.~2 of
\citet{1996ApJ...456..811H} and a temperature of 4000--4500
K\footnote{This excitation temperature for SNe Ib/c at late phases is
  obtained from specific models of nebular spectra
  \protect\citep{2001ApJ...559.1047M,2007ApJ...670..592M,2010MNRAS.408...87M}.},
the flux ratio [\OI] $\lambda$6300\,/$\lambda$5577 of about 20--40 ,
as measured from our TNG spectrum, indicates an electron density of
$\geq$ $10^8$ cm$^{-3}$.  The flux of the [\OI]
$\lambda\lambda6300,6364$ feature measured at 270 days ($3.7 \times
10^{-14}$ erg cm$^{-2}$ s$^{-1}$) can also be used to roughly estimate
the ejected oxygen mass. Using the equation from
\citet{1986ApJ...310L..35U} that is appropriate for our
electron-density and temperature regimes, we obtain an oxygen mass of
$M_\mathrm{O}$\,$\sim$\,5--10 \Mo. As a comparison, the oxygen masses
obtained with the same method for SNe 1990I \citep[0.7--1.35
  \Mo,][]{2004A&A...426..963E} and 2009jf \citep[1.34
  \Mo,][]{2011MNRAS.413.2583S} are significantly smaller. Also the
oxygen mass of SN~1998bw \citep[$\sim$\,3
  \Mo,][]{2001ApJ...559.1047M}, computed via spectral modelling, is a
factor of 2 smaller than that of SN~2011bm.  Such a large oxygen mass
in the ejecta is expected when the progenitor is either a very massive
star ($\geq$ 30 \Mo{}) or a lower-mass star formed in a very low
metallicity environment
\citep{1996ApJ...460..408T,2003ApJ...592..404L,2005A&A...433.1013H,2006NuPhA.777..424N}.
Given the metallicity measurement reported above, it is likely that
the progenitor of SN~2011bm must have been very massive. Since most
explosion models of massive stars predict that the oxygen mass
represents 60 to 70$\%$ of the total ejected mass \citep[see
  e.g.][]{2003ApJ...592..404L,2006NuPhA.777..424N}, we can estimate
the total ejected mass to be in the range 7--17 \Mo{}.

\begin{figure*}
\hspace{0.5in}
\epsscale{1}
\plotone{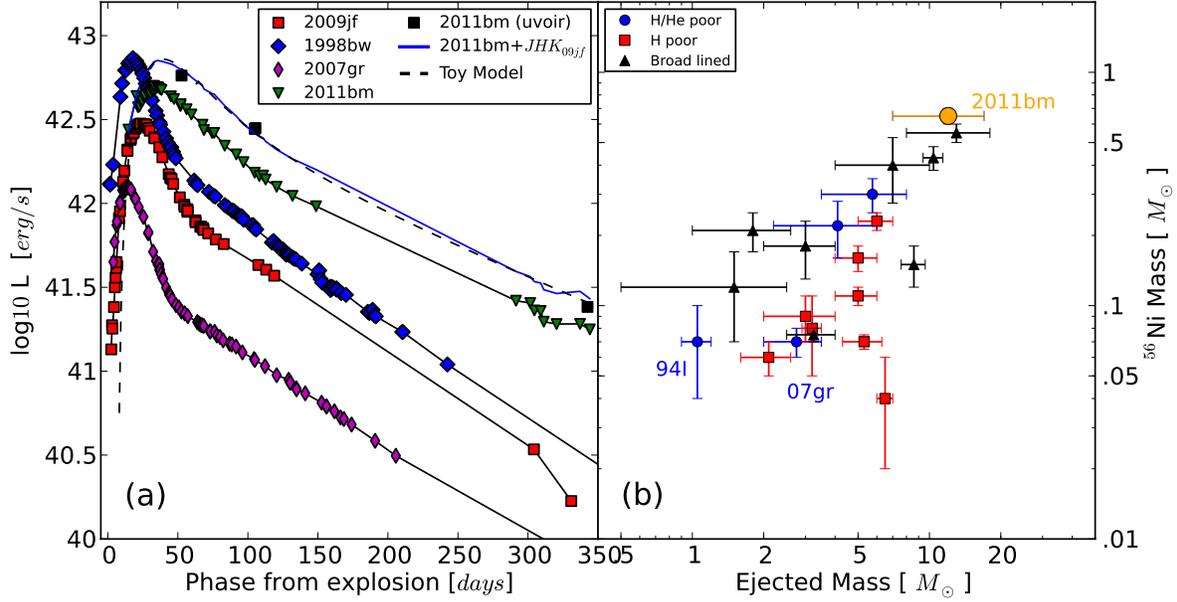}
\caption{\textbf{a}: Pseudo-bolometric (UBVRI) light curve of
  SN~2011bm, compared with a set of pseudo-bolometric light curves of
  other stripped-envelope SNe. The blue solid line is the {\it uvoir}
  bolometric luminosity of SN~2011bm, computed with the UBVRI-light
  curve mentioned before corrected by the fractional NIR contribution
  of SN~2009jf \citep{2011MNRAS.416.3138V}. The resulting {\it uvoir}
  bolometric light curve is consistent with the {\it uvoir} bolometric
  luminosity obtained for the few epochs in which JHK photometry of
  SN~2011bm is available (labelled as $uvoir$). The dashed line
  corresponds to the fit to the computed {\it uvoir} bolometric
  luminosity with the toy model described in
  \protect\citet{2008MNRAS.383.1485V}.  \textbf{b}: Synthesized
  $^{56}$Ni mass versus ejecta mass for a heterogeneous sample of
  core-collapse SNe.  Data are from
  \protect\citet{2009ApJ...692.1131T,2009A&A...508..371H,2011MNRAS.416.3138V,2009ApJ...696..713S}.
}
\label{fig3}
\end{figure*}

In Fig.~\ref{fig3}a we show the quasi-bolometric light curve of
SN~2011bm obtained by integrating our multi-band (UBVRI)
observations. We also show the {\it uvoir} bolometric luminosity
obtained for the few epochs in which JHK photometry of SN~2011bm is
available (labelled as $uvoir$ in Fig. \ref{fig3}a) and a computed
{\it uvoir} bolometric light curve using the fractional NIR
contribution of SN~2009jf \citep{2011MNRAS.416.3138V}.  The late-time
tail of this light curve is flatter than those of other
stripped-envelope SNe, but still steeper than of H-rich core-collapse
SNe powered by $^{56}$Co decay. This is a clear indication that the
$\gamma$-rays produced in the $^{56}$Co decay are not fully
trapped. However, a larger trapping fraction than in other type Ic SNe
confirms that the ejecta of SN~2011bm are quite massive. We used the
toy model described in \cite{2008MNRAS.383.1485V}\footnote{The toy
  model is based on the Arnett's rule
  \protect\citep{1982ApJ...253..785A} for the photospheric phase,
  equations 1 and 2 from \protect\cite{1997A&A...328..203C} and a
  simple $\gamma$-ray deposition function
  \protect\citep{1997ApJ...491..375C}.} to independently estimate the
physical conditions at the moment of the explosion. The $uvoir$
bolometric light curve can be reproduced with an explosion of a
massive star releasing 7-16 foe of kinetic energy in ejecta of 7-10
\Mo{}, 0.6-0.7 \Mo{} of which is synthesized $^{56}$Ni. These values
are in agreement with those given above. The Nickel mass is
surprisingly high, if we consider that SN 2011bm was less luminous
than SN 1998bw at maximum.

\section{Discussion and conclusion}
The extremely slow evolution of the light curve and spectra of
SN~2011bm have never been observed in other \emph{normal}
stripped-envelope SNe. Despite this, its spectral features are very
similar to those observed in typical SNe~Ic, and much narrower than in
broad-lined events. These observables are consistent with the
explosion of a very massive star (with a main-sequence mass of $\geq
30$ \Mo{}) that ejected 7--17 \Mo{} of material, 5--10 \Mo{} of which
is oxygen and 0.6--0.7 \Mo{} is $^{56}$Ni). The kinetic energy lies
between 7 and 17 foe.  SN 2011bm is the first stripped-envelope SN
with such a slow evolution and such peculiar physical parameters. The
intrinsic rareness of these events supports the idea that they arise
from very massive stars. To our knowledge, only one other
\emph{normal} stripped envelope SN has been proposed to come from a
similarly massive star \citep[SN~1984L, ][]{1991ApJ...379L..13S}.

An alternative, more exotic scenario to explain the luminous and broad
light curve of SN~2011bm without invoking exceptionally high $^{56}$Ni
and total ejected masses is that of a magnetar-powered core-collapse
SN. In this case the light curve of SN~2011bm would be dominated by
the energy released in the spin-down of a newly formed magnetar.
However, the slope of the late-time light curve of SN~2011bm (until a
phase of 300 days) is consistent with that expected in a normal
radioactively-powered SN, without the need of an additional source as
proposed in models of magnetar-driven events
\citep{2010ApJ...717..245K,2010ApJ...719L.204W}\footnote {In addition,
  the handful of candidate magnetar-powered SNe recently discovered
  have never shown a late-time light-curve flattening onto the
  $^{56}$Co tail \citep[see
    e.g.][]{2010ApJ...724L..16P,2011Natur.474..487Q}.}.

Another characteristic of a magnetar-powered SN would be a bi-polar
geometry of its ejecta \citep{2010ApJ...717..245K}. Evidence of a
bi-polar explosion may be found in the profiles of the oxygen lines in
nebular spectra. Double-peaked profiles would support the idea of a
bi-polar explosion, whereas the absence of a double peak is not
sufficient to rule out such an explosion geometry. Following the
approach of \citet{2009MNRAS.397..677T}, we inspected the oxygen
feature and, though a weak double peak is marginally visible, it
appears consistent with the doublet nature of [\OI]
$\lambda\lambda6300,6364$.  In conclusion, our observations do neither
require nor support a magnetar-SN scenario. Instead, the traditional
framework of a $^{56}$Ni-powered core-collapse SN from a very massive
star seems to provide a coherent explanation. This makes SN~2011bm the
most massive and $^{56}$Ni-rich \emph{normal} (i.e. not broad-lined)
SN~Ic ever observed. Further analysis and modelling will certainly
help in better understanding this exceptional event.

\acknowledgments S.V. is grateful to H. Wang for hospitality at the
UCLA. The authors thank E.~Gall, R.~Pakmor and I.~Maurer for
assistance with the observations. S.B., E.C., M.T.B. and M.T. are
partially supported by the PRIN-INAF 2009 with the project
\emph{Supernovae Variety and Nucleosynthesis
  Yields}. S.T. acknowledges support by the TRR 33 \emph{The Dark
  Universe} of the German Research Foundation.  The PS1 Surveys have
been made possible through contributions of the Institute for
Astronomy, the University of Hawaii, the Pan-STARRS Project Office,
the Max-Planck Society and its participating institutes, the Max
Planck Institute for Astronomy, Heidelberg and the Max Planck
Institute for Extraterrestrial Physics, Garching, The Johns Hopkins
University, Durham University, the University of Edinburgh, Queen's
University Belfast, the Harvard-Smithsonian Center for Astrophysics,
and the Las Cumbres Observatory Global Telescope Network,
Incorporated, the National Central University of Taiwan, and the
National Aeronautics and Space Administration under Grant
No. NNX08AR22G issued through the Planetary Science Division of the
NASA Science Mission Directorate.

{\it Facilities:} \facility{ESO-NTT (EFOSC2 and SOFI)},
\facility{Liverpool Telescope (RATCAM)}, \facility{1.82m Asiago
  Mt. Ekar Telescope (AFOSC)}, \facility{Calar Alto 2.2m Telescope
  (CAFOS)}, \facility{LBT (LUCIFER)}, \facility{NOT (ALFOSC)},
\facility{PS1}, \facility{TNG (LRS and NICS)}, \facility{Faulkes
  Telescope North (FS02)}, \facility{William Herschel Telescope (ISIS
  and ACAM)}.\\

\clearpage

\begin{deluxetable}{lccccccl}
\tablecaption{Photometric Data$^a$}
\tabletypesize{\scriptsize}
\tablewidth{0pt}
\tablehead{\colhead{Date} & \colhead{JD$-2400000$} & \colhead{u} & \colhead{g} & \colhead{r} & 
\colhead{i} & \colhead{z} & \colhead{Telescope $+$ Instrument}} 
\startdata 
 2011 Mar 26  & 55646.53   &  --           &  --                & 18.71  0.10   & 18.82  0.06   &  --           &  PS1 + GPC1 \\
 2011 Mar 29  & 55649.56   &  --           &  18.78  0.16  &  --                 &   --               &  --           &  PS1 + GPC1 \\
 2011 Apr 11  & 55663.50   & 17.42  0.08   & 17.17  0.16   & 16.95  0.15   & 16.93  0.27   & 17.07  0.12   &  Liverpool Telescope + RATCAM \\
 2011 Apr 12  & 55664.50   & 17.56  0.03   & 17.07  0.19   & 16.84  0.18   & 16.84  0.07   & 16.92  0.05   &  Liverpool Telescope + RATCAM \\
 2011 Apr 13  & 55665.37   &  --           & 16.93  0.42   & 16.81  0.05   & 16.74  0.04   & 16.82  0.09   &  Liverpool Telescope + RATCAM \\
 2011 Apr 14  & 55666.42   & 17.39  0.10   & 16.88  0.28   & 16.69  0.19   & 16.71  0.16   & 16.83  0.06   &  Liverpool Telescope + RATCAM \\
 2011 Apr 20  & 55672.42   & 17.42  0.06   & 16.82  0.06   & 16.54  0.11   & 16.53  0.11   & 16.63  0.03   &  Liverpool Telescope + RATCAM \\
 2011 Apr 23  & 55675.38   &  --           & 16.74  0.09   & 16.48  0.08   & 16.49  0.14   & 16.59  0.02   &  Liverpool Telescope + RATCAM \\
 2011 Apr 24  & 55676.41   & 17.52  0.23   & 16.82  0.04   & 16.43  0.04   & 16.43  0.05   & 16.53  0.03   &  Liverpool Telescope + RATCAM \\
 2011 Apr 25  & 55677.44   & 17.60  0.15   & 16.82  0.09   & 16.42  0.08   & 16.41  0.07   & 16.58  0.03   &  Liverpool Telescope + RATCAM \\
 2011 May 03  & 55685.37   & 18.40  0.10   & 17.02  0.06   & 16.34  0.07   & 16.41  0.05   & 16.51  0.02   &  Liverpool Telescope + RATCAM \\
 2011 May 07  & 55689.60   & 18.71  0.20   & 17.27  0.04   & 16.33  0.04   & 16.39  0.03   & 16.49  0.03   &  Liverpool Telescope + RATCAM \\
 2011 May 11  & 55693.38   & 18.92  0.26   & 17.37  0.03   & 16.41  0.03   & 16.44  0.01   & 16.54  0.02   &  Liverpool Telescope + RATCAM \\
 2011 May 14  & 55696.40   & 19.30  0.27   & 17.42  0.07   & 16.52  0.07   & 16.49  0.10   & 16.60  0.07   &  Liverpool Telescope + RATCAM \\
 2011 May 21  & 55703.45   & 19.50  0.16   & 17.63  0.05   & 16.62  0.05   & 16.58  0.03   & 16.69  0.04   &  Liverpool Telescope + RATCAM \\
 2011 May 27  & 55709.47   & 19.57  0.12   & 17.75  0.08   & 16.81  0.07   & 16.65  0.03   & 16.71  0.03   &  Liverpool Telescope + RATCAM \\
 2011 Jun 04  & 55717.49   & 19.74  0.17   & 17.96  0.04   & 16.92  0.04   & 16.82  0.02   & 16.79  0.03   &  Liverpool Telescope + RATCAM \\
 2011 Jun 16  & 55729.40   & 19.97  0.24   & 18.21  0.08   & 17.22  0.08   & 17.16  0.04   & 16.97  0.05   &  Liverpool Telescope + RATCAM \\
 2011 Jun 26  & 55739.42   & 19.99  0.31   & 18.37  0.06   & 17.35  0.06   & 17.30  0.05   & 17.02  0.03   &  Liverpool Telescope + RATCAM \\
 2011 Jul 07  & 55750.39   & 20.18  0.22   & 18.55  0.06   & 17.62  0.05   & 17.49  0.03   & 17.19  0.05   &  Liverpool Telescope + RATCAM \\
 2011 Jul 17  & 55760.44   & 20.22  0.65   & 18.71  0.18   & 17.74  0.14   & 17.64  0.08   & 17.39  0.05   &  Liverpool Telescope + RATCAM \\
 2011 Dec 14  & 55909.69   &  --           & 20.20  0.61   & 19.43  0.36   & 19.46  0.31   & 19.67  0.98   &  Liverpool Telescope + RATCAM \\
 2011 Dec 31  & 55927.77   &  --           & 20.47  0.27   & 19.51  0.26   & 19.65  0.08   & 19.77  0.16   &  Liverpool Telescope + RATCAM \\
\hline\\[-0.2cm]
\quad\ Date & JD$-2400000$ & U & B & V & R & I & \hspace*{1.5cm} Telescope \\[0.1cm]
\hline\\[-0.2cm]  
 2011 Apr 10  & 55662.56   &  --           & 17.26  0.11   & 16.91  0.11   & 16.67  0.10   & 16.61  0.06   &  1.82m Ekar Telescope + AFOSC \\
 2011 Apr 11  & 55663.49   & 16.80  0.30   & 17.31  0.04   & 16.85  0.11   & 16.56  0.11   & 16.60  0.11   &  Liverpool Telescope + RATCAM \\
 2011 Apr 12  & 55664.50   & 16.72  0.09   & 17.25  0.04   & 16.83  0.04   & 16.52  0.08   & 16.42  0.09   &  Liverpool Telescope + RATCAM \\
 2011 Apr 13  & 55665.36   &  --           & 17.23  0.10   & 16.82  0.04   & 16.51  0.06   & 16.44  0.05   &  Liverpool Telescope + RATCAM \\
 2011 Apr 14  & 55666.41   & 16.63  0.11   & 17.12  0.05   & 16.80  0.05   & 16.48  0.03   & 16.39  0.02   &  Liverpool Telescope + RATCAM \\
 2011 Apr 14  & 55666.87   &  --           & 17.10  0.07   & 16.72  0.06   & 16.45  0.07   & 16.37  0.04   &  Faulkes Telescope North + EM01 \\
 2011 Apr 16  & 55668.86   &  --           & 17.07  0.11   & 16.70  0.09   & 16.44  0.11   & 16.24  0.09   &  Faulkes Telescope North + EM01 \\
 2011 Apr 17  & 55669.80   & 16.70  0.38   & 17.05  0.08   & 16.67  0.08   & 16.39  0.08   & 16.25  0.08   &  Faulkes Telescope North + EM01 \\
 2011 Apr 18  & 55670.45   & 16.68  0.08   & 17.00  0.08   & 16.71  0.08   & 16.42  0.08   & 16.37  0.08   &  Calar Alto 2.2m Tel. + CAFOS \\
 2011 Apr 19  & 55671.58   & 16.70  0.11   & 17.02  0.12   & 16.65  0.10   & 16.37  0.17   & 16.17  0.15   &  Calar Alto 2.2m Tel. + CAFOS \\
 2011 Apr 19  & 55671.80   & 16.61  0.06   & 17.05  0.15   & 16.61  0.15   & 16.32  0.09   & 16.18  0.08   &  Faulkes Telescope North + EM01 \\
 2011 Apr 20  & 55672.40   & 16.64  0.07   & 17.02  0.05   & 16.61  0.05   & 16.35  0.04   & 16.24  0.07   &  Calar Alto 2.2m Tel. + CAFOS \\
 2011 Apr 20  & 55672.42   & 16.56  0.19   & 17.00  0.19   & 16.58  0.20   & 16.22  0.12   & 16.15  0.11   &  Liverpool Telescope + RATCAM \\
 2011 Apr 20  & 55672.90   & 16.60  0.18   & 17.02  0.06   & 16.54  0.03   & 16.23  0.07   & 16.14  0.10   &  Faulkes Telescope North + EM01 \\
 2011 Apr 23  & 55674.89   &  --           & 17.02  0.07   & 16.53  0.07   & 16.24  0.03   & 16.02  0.04   &  Faulkes Telescope North + EM01 \\
 2011 Apr 23  & 55675.38   &  --           & 17.04  0.09   & 16.54  0.09   & 16.20  0.07   & 16.04  0.06   &  Liverpool Telescope + RATCAM \\
 2011 Apr 24  & 55676.41   & 16.89  0.10   & 17.10  0.06   & 16.54  0.05   & 16.17  0.05   & 16.04  0.05   &  Liverpool Telescope + RATCAM \\
 2011 Apr 25  & 55677.44   & 17.06  0.05   & 17.11  0.02   & 16.52  0.02   & 16.17  0.04   & 16.00  0.03   &  Liverpool Telescope + RATCAM \\
 2011 May 01  & 55683.41   &  --           & 17.32  0.09   & 16.58  0.17   & 16.15  0.06   & 15.95  0.07   &  1.82m Ekar Telescope + AFOSC \\
 2011 May 03  & 55685.37   & 17.56  0.08   & 17.41  0.27   & 16.61  0.20   & 16.14  0.06   & 15.96  0.05   &  Liverpool Telescope + RATCAM \\
 2011 May 07  & 55689.49   & 17.71  0.17   & 17.58  0.16   & 16.63  0.14   & 16.24  0.14   & 16.12  0.09   &  Calar Alto 2.2m Tel. + CAFOS \\
 2011 May 07  & 55689.60   & 17.83  0.04   & 17.61  0.05   & 16.65  0.05   & 16.15  0.05   & 15.99  0.04   &  Liverpool Telescope + RATCAM \\
 2011 May 11  & 55693.37   & 18.06  0.17   & 17.70  0.03   & 16.75  0.03   & 16.20  0.04   & 15.99  0.04   &  Liverpool Telescope + RATCAM \\
 2011 May 14  & 55696.39   & 18.31  0.06   & 17.80  0.07   & 16.85  0.07   & 16.24  0.05   & 16.02  0.04   &  Liverpool Telescope + RATCAM \\
 2011 May 21  & 55703.45   & 18.68  0.12   & 18.18  0.03   & 16.99  0.03   & 16.36  0.03   & 16.02  0.02   &  Liverpool Telescope + RATCAM \\
 2011 May 22  & 55704.36   & 18.52  0.14   & 18.35  0.13   & 16.99  0.13   & 16.40  0.09   & 16.26  0.09   &  Calar Alto 2.2m Tel. + CAFOS \\
 2011 May 27  & 55709.47   & 18.76  0.07   & 18.29  0.04   & 17.16  0.04   & 16.46  0.04   & 16.15  0.04   &  Liverpool Telescope + RATCAM \\
 2011 May 29  & 55710.77   &  --           & 18.30  0.10   & 17.09  0.10   & 16.46  0.07   & 16.20  0.03   &  Faulkes Telescope North + EM01 \\
 2011 Jun 04  & 55717.49   & 18.96  0.22   & 18.49  0.06   & 17.35  0.06   & 16.65  0.03   & 16.32  0.03   &  Liverpool Telescope + RATCAM \\
 2011 Jun 11  & 55724.81   &  --           & 18.60  0.09   & 17.44  0.08   & 16.83  0.05   & 16.45  0.06   &  Faulkes Telescope North + EM01 \\
 2011 Jun 16  & 55729.39   & 19.05  0.36   & 18.76  0.14   & 17.55  0.14   & 16.94  0.03   & 16.58  0.02   &  Liverpool Telescope + RATCAM \\
 2011 Jun 24  & 55736.50   & 19.24  0.07   & 18.89  0.06   & 17.63  0.05   & 17.15  0.04   & 16.75  0.04   &  NTT + EFOSC2 \\
 2011 Jun 26  & 55739.42   & 19.30  0.16   & 18.92  0.09   & 17.69  0.09   & 17.21  0.10   & 16.80  0.09   &  Liverpool Telescope + RATCAM \\
 2011 Jun 28  & 55741.77   &  --           & 18.82  0.10   & 17.69  0.10   & 17.24  0.11   & 16.78  0.04   &  Faulkes Telescope North + EM01 \\
 2011 Jun 30  & 55743.39   & 19.42  0.37   & 18.92  0.17   & 17.80  0.16   & 17.26  0.11   & 16.82  0.09   &  Calar Alto 2.2m Tel. + CAFOS \\
 2011 Jul 07  & 55750.39   & 19.40  0.17   & 19.06  0.07   & 17.90  0.06   & 17.33  0.10   & 16.99  0.09   &  Liverpool Telescope + RATCAM \\
 2011 Jul 17  & 55760.44   & 19.32  0.33   & 19.24  0.08   & 18.04  0.08   & 17.45  0.06   & 17.17  0.06   &  Liverpool Telescope + RATCAM \\
 2011 Aug 01  & 55775.38   &  --           & 19.40  0.22   & 18.23  0.08   & 17.68  0.09   & 17.18  0.08   &  TNG + LRS \\
 2011 Dec 06  & 55902.07   &  --           & 20.70  0.23   & 19.63  0.23   & 19.08  0.27   & 18.71  0.17   &  Faulkes Telescope North + EM01 \\
 2011 Dec 14  & 55911.11   &  --           & 20.92  0.61   & 19.80  0.60   & 18.97  0.34   & 18.61  0.23   &  Faulkes Telescope North + EM01 \\
 2011 Dec 17  & 55913.72   &  --           & 20.75  0.26   & 20.00  0.25   & 19.13  0.22   & 18.84  0.09   &  TNG + LRS \\
 2011 Dec 21  & 55917.67   &  --           & 20.86  0.62   & 19.88  0.50   & 19.16  0.51   & 18.85  0.43   &  WHT + ISIS \\
 2011 Dec 23  & 55919.67   &  --           &  --                & 20.05  0.19   & 19.36  0.19   & 19.09  0.39   &  Calar Alto 2.2m Tel. + CAFOS \\
 2011 Dec 31  & 55927.76   &  --           &  --                & 20.15  0.17   & 19.36  0.23   & 19.10  0.21   &  Liverpool Telescope + RATCAM \\
 2012 Jan 19  &  55948.83   &  --           &  21.07 0.41  &   20.35 0.12   &  19.38  0.12  &  19.08 0.06  &  NTT + EFOSC2 \\
\hline\\[-0.2cm]
\quad\ Date & JD$-2400000$ & J & H & K & -- & -- & \hspace*{1.5cm} Telescope \\[0.1cm]
\hline\\[-0.2cm]
 2011 May 09  & 55690.13   & 15.78  0.16   & 15.58  0.16   & 15.32  0.11   &  --           &  --           &  NTT + SOFI \\
 2011 Jun 24  & 55736.97   & 16.23  0.18   & 15.90  0.18   & 16.05  0.17   &  --           &  --           &  NTT + SOFI \\
 2011 Jun 24  & 55736.50   &  --              & 16.11  0.10   &  --           &  --           &  --           &  LBT + LUCIFER \\
 2012 Jan 21  &  55947.37  &  19.91 0.25 & 19.167 0.25  &   --         &   --           &  --           &  NTT+SOFI
\enddata
\tablenotetext{a}{ugriz magnitudes are calibrated to the Sloan
  photometric system (AB system) using the SDSS catalogue
  \protect\citep{2009ApJS..182..543A}; UBVRI magnitudes are in the
  Bessell system, calibrated with Landolt fields observed during
  photometric nights
  \protect\citep{1992AJ....104..340L}. Near-Infrared images are
  calibrated to 2MASS catalogue \protect\citep{2006AJ....131.1163S}}
\label{tab1}
\end{deluxetable}

\begin{deluxetable}{ccccccc}
\tabletypesize{\scriptsize}
\tablecaption{Spectra collected}
\tablewidth{0pt}
\tablehead{\colhead{Date} & \colhead{JD$-2400000$} & \colhead{Phase}$\!\!^{a}$ & \colhead{Telescope} & \colhead{Instrument} & \colhead{Range [\AA]} &\colhead{Resolution [\AA]}}
\startdata
2011 Apr 11  & 55662.54 & 18.5   &  1.82m Ekar       &  AFOSC+GR04       &  3800--8200 &  20   \\
2011 Apr 14  & 55665.55 & 21.5   &  NOT              &  ALFOSC+grism-4   &  3250--9100 &  14   \\
2011 Apr 18  & 55670.48 & 26.5   &  2.2m Calar Alto  &  CAFOS+blue-200   &  3400--8900 &  14   \\
2011 Apr 20  & 55672.43 & 28.4   &  2.2m Calar Alto  &  CAFOS+blue-200   &  3400--8900 &  14   \\
2011 Apr 24  & 55675.62 & 31.6   &  NOT              &  ALFOSC+grism-4   &  3300--9100 &  14   \\
2011 May 08  & 55690.12 & 46.1   &  2.2m Calar Alto  &  CAFOS+green-200  & 3790--10000 &  14   \\
2011 May 15  & 55697.49 & 53.5   &  TNG              &  LRS+LRB/LRR      & 3200--10000 &  13   \\
2011 May 22  & 55704.39 & 60.4   &  2.2m Calar Alto  &  CAFOS+green-200  & 4800--10200 &  13   \\
2011 Jun 12  & 55724.58 & 80.6   &  TNG              &  LRS+LRB          &  3200--8700 &  12   \\
2011 Jun 24  & 55736.52 & 92.5   &  NTT              &  EFOSC2+Gr13      &  3700--9250 &  18   \\
2011 Aug 10  & 55784.36 & 140.4  &  2.2m Calar Alto  &  CAFOS+green-200  &  3900--9300 &  14   \\
2011 Aug 11  & 55785.37 & 141.4  &  2.2m Calar Alto  &  CAFOS+green-200  &  3900--9300 &  14   \\
2011 Dec 18  & 55913.75 & 269.7  &  TNG              &  LRS+LRR          & 5000--10100 &  10   \\
2011 Dec 20  & 55915.77 & 271.8  &  WHT              &  ISIS+R158R       & 3300--10000 &  6/10 \\
2012 Jan 22  & 55948.86 & 304.9  &  NTT              &  EFOSC2+Gr13      &  3700--9250 &  18 \\
\hline\\[-0.2cm]
2011 May 05 & 55686.62 & 42.7  & TNG                &  NICS +IJ          &   9000--15000 &  20     \\
2011 May 09 & 55690.69 & 46.7  &  NTT                &  SOFI+GB/GR   &   9000--25000 &  20/30 \\
2011 Jun  24 & 55737.50 & 93.5  &  NTT                &  SOFI+GB         &   9000--15000 &  20       
\enddata
\tablenotetext{a}{Phase with respect to the estimated explosion epoch.}
\label{tab2}
\end{deluxetable}
\end{document}